\documentclass[aps,pre,onecolumn,superscriptaddress,longbibliography,notitlepage,10pt]{revtex4-1}
\usepackage{mathtools,amsmath,amssymb}
\usepackage{graphicx}
\usepackage{color}
\usepackage{bm}

\newcommand{\bs}{\boldsymbol}

\begin{document}

\title{Turbulence and turbulent pattern formation in a minimal model for active fluids}

	\author{Martin James}
	\affiliation{Max Planck Institute for Dynamics and Self-Organization (MPI DS), Am Fa{\ss}berg 17, 37077 G\"{o}ttingen, Germany}
	\author{Wouter J.T. Bos}
	\affiliation{LMFA, CNRS, \'Ecole Centrale de Lyon, Universit\'e de Lyon, 69134 Ecully, France}
	\author{Michael Wilczek}
	\email{michael.wilczek@ds.mpg.de}
	\affiliation{Max Planck Institute for Dynamics and Self-Organization (MPI DS), Am Fa{\ss}berg 17, 37077 G\"{o}ttingen, Germany}
				
\date{August 10, 2018}

\begin{abstract}
Active matter systems display a fascinating range of dynamical states, including stationary patterns and turbulent phases. While the former can be tackled with methods from the field of pattern formation, the spatio-temporal disorder of the active turbulence phase calls for a statistical description. Borrowing techniques from turbulence theory, we here establish a quantitative description of correlation functions and spectra of a minimal continuum model for active turbulence. Further exploring the parameter space, we also report on a surprising type of turbulence-driven pattern formation far beyond linear onset: the emergence of a dynamic hexagonal vortex lattice state after an extended turbulent transient, which can only be explained taking into account turbulent energy transfer across scales.
\end{abstract}


\maketitle

\section{INTRODUCTION}

Flows driven by active agents display a rich variety of dynamical states \cite{ramaswamy10arcm,marchetti13rmp,yeomans2014natmat}. Active stresses and hydrodynamics collude to create collective motion, both regular and chaotic, in systems of motile micro-organisms \cite{mendelson99jba,dombrowski04prl,ishikawa08prl} or artificial self-propelled agents \cite{howse07prl,bricard13nat} on scales much larger than the individual. For example, sufficiently dense suspensions of motile micro-organisms, such as {\it B. Subtilis}, exhibit a spatio-temporally disordered phase. Owing to its reminiscence of hydrodynamic turbulence, this phenomenon has been termed active turbulence \cite{wolgemuth08bpj,wensink12pnas,dunkel13prl,bratanov15pnas,thampi16epj,genkin17prx}. Similar observations were also reported in systems dominated by nematic interactions such as ATP-driven microtubule networks \cite{sanchez12nature}.  Besides active turbulence, remarkably ordered phases were found in a number of systems. Self-organized vortex lattices, for example, have been discovered both in hydrodynamically interacting systems, such as spermatozoa \cite{riedel05science}, as well as in dry microtubule systems \cite{sumino2012nature}. Confinement offers yet another possibility of organizing flows into regular large-scale flow \cite{suzuki17pnas} and vortex patterns \cite{wioland16nap}.
 
The occurrence of these phenomena in vastly different systems has motivated the development and exploration of a range of minimal mathematical models. They can be broadly categorized into agent-based models of self-propelled particles with nematic or polar interactions \cite{vicsek95prl,marchetti13rmp,grossmann14prl,grossmann15epje,bechinger16rmp} and continuum theories for a small number of order parameters \cite{wolgemuth08bpj,wensink12pnas,dunkel13njp,giomi15prx,urzay17jfm}. These models have been shown to capture a variety of dynamical phases of active fluids, including active turbulence and vortex lattice states. For example, in \cite{wensink12pnas} the active turbulence phase was modeled and compared with experiments. Regarding ordered phases, vortex lattices have been observed and investigated at the crossover from the hydrodynamic to the friction-dominated regimes of models for confined active fluids \cite{doostmohammadi16natcomm}. These systems display phases of two-signed vortices with length scales defined by the dimensions of the system. In a class of particle-based models for active matter, the emergence of vortex lattices has been related to a classical pattern formation mechanism as a result of a Turing instability \cite{grossmann14prl,grossmann15epje}.

While many such models have been shown to capture the dynamics of active systems qualitatively and quantitatively, the complexity of disordered states like active turbulence eventually calls for a statistical description. The goal of such a non-equilibrium statistical mechanics of active matter is the computation of fundamental statistical quantities such as correlation functions without resorting to expensive numerical integration of systems with thousands or even millions of degrees of freedom.

Recent developments of statistical theories on top of minimal continuum theories for active matter have provided insights into the small-scale correlation structure of an active nematic fluid based on a mean field approach for the vorticity field \cite{giomi15prx}, as well as a theory capturing large-scale features of polar bacterial flows based on analytical closure techniques \cite{bratanov15pnas}. A theoretical framework capturing the correlation function or equivalently the spectral properties for the full range of scales of such prototypical active systems, however, is currently lacking.

In this Rapid Communication, we set out to close this gap. Borrowing techniques from turbulence theory, we derive correlation functions and spectra of the turbulent phase of the minimal continuum theory recently established in \cite{wensink12pnas} to capture the dynamics of dense bacterial suspensions. Further exploring the parameter space, we also discover a novel phase of turbulent pattern formation, i.e.~an extensive turbulent transient governed by strong advection which eventually results in a highly ordered vortex lattice state. We demonstrate that turbulence characteristics crucially contribute to the emergence of this novel pattern through nonlinear advective energy transfer. This mechanism differs profoundly from the classical route to pattern formation. To make this transparent, we first briefly recapitulate classical pattern formation in this minimal model for active fluids in absence of nonlinear advection.

\subsection{Minimal Model for Active Fluids}

The starting point is the equation for active turbulence as proposed in \cite{wensink12pnas,dunkel13njp} for a two-dimensional incompressible velocity field $\bs u(\bs x,t)$ describing the coarse-grained dynamics of a dense bacterial suspension. It takes the nondimensionalized form \footnote{For the nondimensionalization we start from the equation presented in \cite{wensink12pnas} and note that the term involving $\lambda_1$ can be absorbed into the pressure gradient term. Then we define the time scale $T = 4 \Gamma_2 / \Gamma_0^2$ and the length scale $L = \sqrt{-2\Gamma_2/\Gamma_0}$ to nondimensionalize the equation. To obtain Eq.~\eqref{eq:equationofmotion}, the parameters in the dimensional equation are mapped to the ones in the nondimensional equation according to $\lambda_0 \rightarrow \lambda$, $\Gamma_0T/L^2 \rightarrow -2$, $\Gamma_2T/L^4 \rightarrow 1$, $\alpha T \rightarrow \alpha+1$ and $\beta L^2/T \rightarrow \beta$. We note that one additional parameter can be scaled out \cite{oza16epje}, which we refrain from here for presentation purposes.}
\begin{equation}
\label{eq:equationofmotion}
  \partial_t \bs u + \lambda \bs u \cdot \nabla \bs u = -\nabla p - (1+\Delta)^2 \bs u - \alpha \bs u - \beta \bs u^2\, \bs u
\end{equation}
and represents a minimal field theory for a polar order parameter field, combining Navier-Stokes dynamics (advective nonlinearity and nonlocal pressure gradient) with elements of pattern forming systems (linear wave number selection and a saturating higher-order nonlinearity). Owing to its similarity to the Navier-Stokes equation, this minimal model is particularly suited to develop a statistical theory with methods from turbulence theory.
\begin{figure}
  \includegraphics[width=1.0\textwidth]{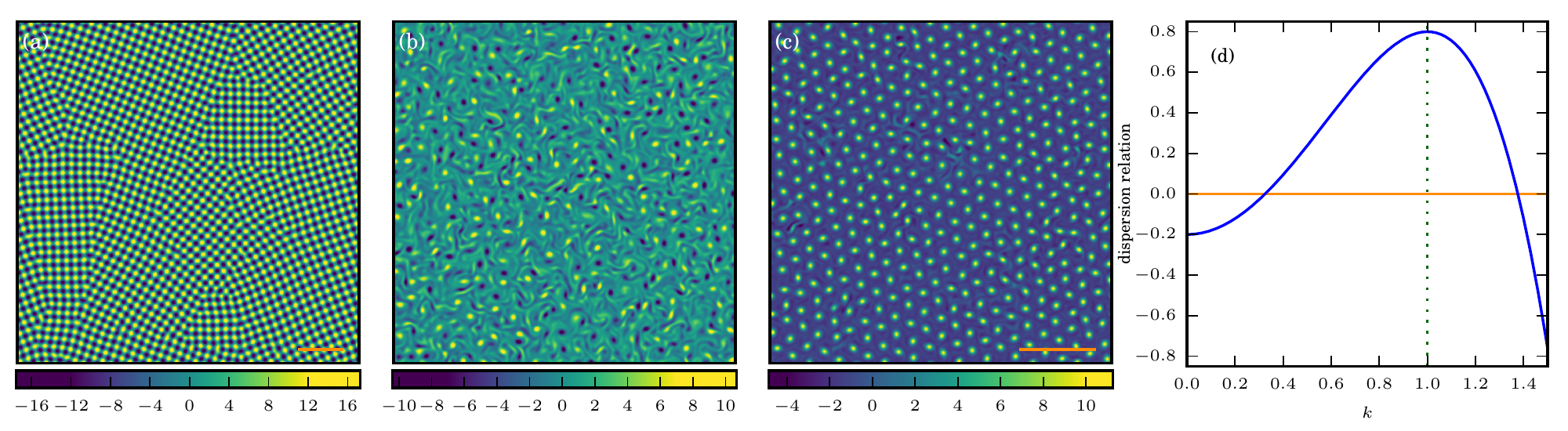}
  \caption{The continuum model Eq.~\eqref{eq:equationofmotion} displays a range of dynamical phases of the vorticity field depending on the nonlinear advection: (a) classical pattern formation ($\lambda=0$, simulation 1 in Table \ref{tab:simpara}), (b) active turbulence ($\lambda=3.5$, simulation 2 in Table \ref{tab:simpara}) and (c) turbulent pattern formation ($\lambda=7$, simulation 3 in Table \ref{tab:simpara}). Notably, the dispersion relation shown in (d) along with the nonlinear damping is kept fixed for all examples. The dashed green line corresponds to the most unstable wave number, given by $k=k_c$, which sets the wave number of the pattern in (a). The horizontal orange lines in (a) and (c) correspond to five times the length scale of the patterns, i.e.~$10\pi$/$k_c$ and $10\pi/k_0$, respectively, exemplifying that the wave number selection in the turbulent pattern forming phase (c) differs from the classical pattern forming phase (a).}
  \label{fig:dynamicalstates}
\end{figure}

\begin{table}
\begin{tabular}{ cccccccc}
  No. & dynamical state & $\lambda$ & $\alpha$ & $\beta$ & $N$ & $D$ & $\Delta t$ \\
  \hline
  1 & square lattice & 0 & -0.8 & 0.01 & 2048 & 250 & $10^{-2}$\\
  2 & active turbulence & 3.5 & -0.8 & 0.01 & 2048 & 250 & $10^{-3}$ \\
  3 & hexagonal lattice & 7.0 & -0.8 & 0.01 & 2048 & 250 & $10^{-3}$ \\
  4 & hexagonal lattice & 7.0 & -0.8 & 0.01 & 2048 & 125 & $10^{-3}$ \\
  5 & active turbulence & 3.5 & -0.3 & 0.01 & 2048 & 250 & $10^{-3}$ \\
  6 & benchmark case \cite{wensink12pnas,bratanov15pnas} & 3.5 & -1.178 & 0.01125 & 2048 & 250 & $10^{-3}$ \\
\end{tabular}
\caption{Simulation parameters. The active fluid is characterized through the parameters $\lambda$, $\alpha$ and $\beta$. The simulations are run on grids with $N^2$ grid points, discretizing a domain of lateral extent $D$; $\Delta t$ denotes the time step.}
\label{tab:simpara}
\end{table}

The dynamical phases of this continuum theory are explored in Fig.~\ref{fig:dynamicalstates}. Unless otherwise noted, we fix $\alpha=-0.8$ and $\beta=0.01$ to focus on the role of nonlinear advection. The results are obtained numerically with a pseudo-spectral code using a second-order Runge-Kutta scheme, and an integrating factor is used for treating the linear terms. More details on the simulations are provided in the supporting information. Table \ref{tab:simpara} lists the range of parameters explored in this manuscript.

\subsection{Classical Pattern Formation}

 For  $\lambda=0$ the equation reduces to a vectorial Swift-Hohenberg type system which follows a gradient dynamics as discussed in the supporting information. In this parameter regime, we observe the emergence of stationary square lattices consistent with previous literature~\cite{dunkel13njp,oza16epje}. Figure~\ref{fig:dynamicalstates}(a) shows a non-ideal square lattice with defects such as grain boundaries from our numerical simulations. As expected, the emergence of this state can be explained with tools from classical pattern formation theory in terms of amplitude equations. We analyze the corresponding amplitude equations \cite{cross09book} of the vorticity formulation of Eq.~\eqref{eq:equationofmotion}. The analysis detailed in the SI reveals the stability of the square lattice state with amplitude $A=\sqrt{-\alpha k_c^2/(5\beta)}$, which corresponds to a maximum value of the field of $4A$. In comparison, single-stripe patterns are linearly unstable. For the investigated parameters given in Table \ref{tab:simpara} the value of the theoretically predicted amplitude is $4.00$, which is confirmed by our simulations to within $5$ percent. This brief exposition serves to show that the classical pattern formation in absence of nonlinear advection leads to a stationary square lattice state with wave number $k_c=1$.

\section{Active Turbulence} 

As the advective term is switched on by setting $\lambda=3.5$, the nonlinear energy transfer sets in, which by generating vortices of larger size renders the stationary square lattice pattern unstable. As a result, a self-sustained turbulence-like phase emerges (see Fig.~\ref{fig:dynamicalstates}(b)), which has been characterized, e.g. in \cite{wensink12pnas,bratanov15pnas,james2018vortex}. Borrowing techniques from classical turbulence theory, we here establish a statistical description for the two-point correlation function and energy spectra for the full range of dynamically active scales.

To this end, we consider the velocity covariance tensor $R_{ij}(\bs r) = \langle u_i(\bs x,t) u_j(\bs x+\bs r,t) \rangle \equiv \langle u_i u_j' \rangle$ which is among the most fundamental statistical objects of interest; by virtue of kinematic relations, it contains the correlation structure of the velocity field as well as of the vorticity and velocity gradient tensor fields \cite{batchelor53book}. Its evolution equation for the statistically homogeneous and isotropic turbulent phase is readily obtained as
\begin{align}\label{eq:covarianceevo}
  &\partial_t R_{ij} + \lambda\partial_k \langle u_k' u_i u_j' - u_k u_i u_j' \rangle = -2\left[ (1+\Delta)^2 + \alpha\right]R_{ij} -\beta \langle u_k u_k u_i u_j' + u_k' u_k' u_i u_j' \rangle \, . 
\end{align}
As a result of statistical isotropy, the pressure contribution vanishes. The quadratic and cubic nonlinearities result in unclosed terms which obstruct a direct computation of the covariance without making further assumptions. The main effect of the $\beta$-term is to saturate the velocity growth. Owing to the approximate Gaussianity of the velocity field \cite{wensink12pnas,dunkel13prl,bratanov15pnas,james2018vortex}, the correlator in this term can be factorized using Wick's theorem, which yields $\langle u_k u_k u_i u_j' + u_k' u_k' u_i u_j' \rangle = 2 R_{kk}(\bs 0)R_{ij}(\bs r) + 2 R_{ik}(\bs 0)R_{kj}(\bs r) + 2 R_{ik}(\bs r) R_{kj}(\bs 0)$.

\begin{figure*}
  \includegraphics[width=1.0\textwidth]{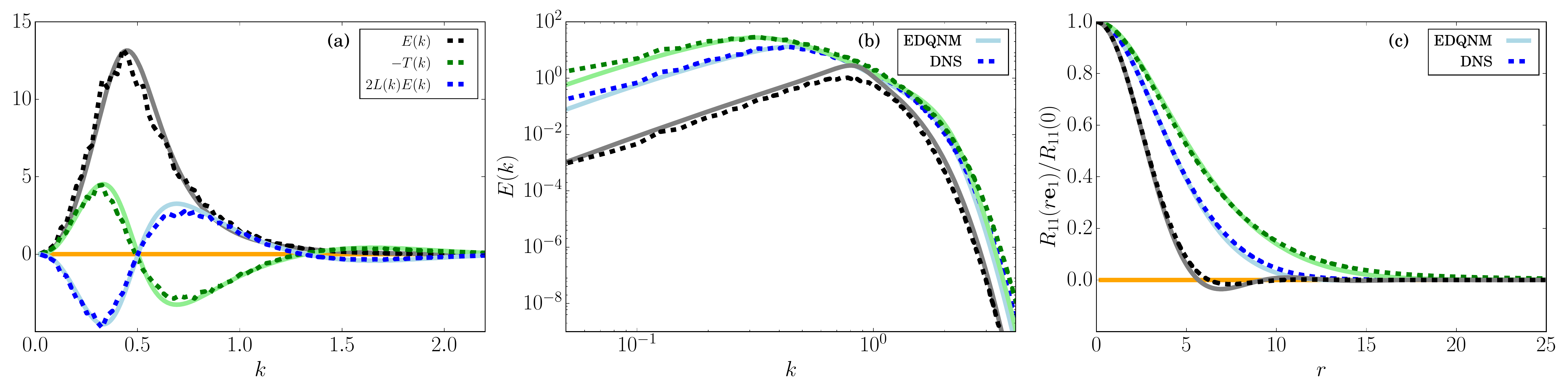}
  \caption{(a) Energy budget of active turbulence: direct numerical simulation (DNS) results (dashed lines, simulation 2 in Table \ref{tab:simpara}) vs EDQNM closure theory. The black, green and blue curves correspond to the energy spectrum, the transfer term and the effective linear term, respectively. (b) Spectra from DNS of active turbulence compared to EDQNM closure theory. (c) Longitudinal velocity autocorrelation of active turbulence: DNS vs EDQNM closure theory. The blue, black and green curves in (b) and (c) correspond to the simulations 2, 5 and 6, respectively, as listed in Table \ref{tab:simpara}.
 }
  \label{fig:edqnmresults}
\end{figure*}

An analogous attempt to factorize the triple correlators fails as this amounts to neglecting the energy transfer across scales, a hallmark feature of turbulence \cite{monin13book}. A more sophisticated closure needs to be established. For the subsequent treatment we choose a Fourier representation of the covariance tensor $R_{ij}(\bs r)$ in terms of the spectral energy tensor $\Phi_{ij}(\bs k)$. For a statistically isotropic two-dimensional flow, it takes the form $\Phi_{ij}(\bs k,t) = E(k,t)/(\pi k)\left[ \delta_{ij} - k_ik_j/k^2 \right]$, where $E(k,t)$ denotes the energy spectrum function. Starting from Eq.~\eqref{eq:covarianceevo}, an evolution equation for the energy spectrum function can be derived which takes the form \cite{batchelor53book,monin13book,pope00book}
\begin{equation}\label{eq:spectrumevo}
  \partial_t E(k,t) + T(k,t) =  2 L(k,t) E(k,t) \, .
\end{equation}
Here, $T(k,t)$ is the energy transfer term between different scales which results from the triple correlators in Eq.~\eqref{eq:covarianceevo}; $L(k,t) = -(1-k^2)^2 - \alpha - 4 \beta E_0(t)$ is the \emph{effective linear term}, which represents all linear terms as well as the Gaussian factorization of the cubic nonlinearity with $E_0(t) = \int E(k,t) \, \mathrm{d}k$. The effective linear term is responsible for the energy injection around $k_c=1$ as well as for the damping at small and large scales. 
For the energy transfer term, we adopt the so-called eddy-damped quasi-normal Markovian (EDQNM) approximation and present here the main steps of the derivation for active fluids. More details are given in the SI. For a more comprehensive account of this model, which has been successfully applied to hydrodynamic turbulence, we refer the reader to \cite{Orszag1974,lesieur2012turbulence,SagautBook}. The core idea of this closure scheme is to consider the evolution equation for the triple correlators in addition to Eq.~\eqref{eq:spectrumevo}, from which $T(k,t)$ can be obtained straightforwardly. The occurring fourth-order moments are then factorized assuming Gaussianity, similar to the treatment of the nonlinear damping term in Eq.~\eqref{eq:covarianceevo}, i.e. $\langle\hat{u}\hat{u}\hat{u}\hat{u}\rangle = \Sigma\langle\hat{u}\hat{u}\rangle\langle\hat{u}\hat{u}\rangle$ (written in a symbolic fashion). The influence of the neglected cumulants is modeled by an additional damping, which leads to an effective damping $\eta_{kpq}$ (see SI for more information). As a result we obtain an evolution equation for the triple correlators of the velocity modes $\bs k$, $\bs p$ and $\bs q$:
\begin{align}
  \left[\partial_t+ \eta_{kpq}\right]\langle \hat{u}({\bs k})\hat{u}({\bs p})\hat{u}({\bs q}) \rangle= \lambda\Sigma\langle\hat{u}\hat{u}\rangle\langle\hat{u}\hat{u}\rangle.
\end{align}
As a next step, we apply the so-called Markovianization by assuming that the right-hand side evolves slowly, such that this equation can be integrated analytically and the steady state solution can be obtained by taking $t\rightarrow \infty$. The energy transfer function, which is a contraction of the triple velocity tensor, can then be written as
\begin{align}\label{eq:edqnm}
T(k,t)=\iint_{\Delta}\frac{\lambda^2}{\eta_{kpq}} \, \big[a(k,p,q)E(p,t)E(q,t)+b(k,p,q)E(q,t)E(k,t)\big]\mathrm{d}p\mathrm{d}q \, .
\end{align}
Here $1/\eta_{kpq}$ acts as a characteristic time scale which results from the turbulent damping. The geometric factors $a(k,p,q)$ and $b(k,p,q)$ are associated to contractions of the isotropic tensor $\langle \hat{u}({\bs k})\hat{u}({\bs p})\hat{u}({\bs q}) \rangle$; the exact expressions of the terms are given in the SI. $\Delta$ restricts the integration domain in $p,q$-space so that the three wave numbers $k,p,q$ form the sides of a triangle. These triadic interactions are a direct consequence of the quadratic advective nonlinearity. While technically quite involved, the key feature is that the energy transfer term is expressed in terms of the energy spectrum only, i.e.~we have obtained a closure. To illustrate the results, the left panel of Fig.~\ref{fig:edqnmresults} shows a comparison of the terms of Eq.~\eqref{eq:spectrumevo} obtained from the EDQNM closure with a direct estimation from simulation data for active turbulence. Very good agreement is found for all wave numbers. Consistent with the observations in \cite{bratanov15pnas}, the energy transfer term takes energy from the linear injection scale and transports it upscale. This inverse energy transfer is typical for two-dimensional flows \cite{davidson15book}. Interpreting these results in the context of bacterial turbulence, the dominant energy injection occurs on a length scale comparable to the individual bacteria \cite{wensink12pnas}, yet their collective motion displays much larger scales. In the framework of the continuum model Eq.~\eqref{eq:equationofmotion}, this collective behavior is the result of an energy transfer to larger scales induced by nonlinear advection. The EDQNM theory captures this effect accurately. Also the effective linear term, which injects energy in a wave number band around $k_c=1$, but extracts energy at large and small scales, is captured accurately, demonstrating the fidelity of the Gaussian factorization of nonlinear damping. The spectra resulting from the EDQNM closure are shown in the middle panel of Fig.~\ref{fig:edqnmresults}. To demonstrate the validity of the closure theory for a broader parameter range, we additionally varied the $\alpha$ parameter (see Table \ref{tab:simpara}). Furthermore, we also compare with the reference case reported in \cite{bratanov15pnas,wensink12pnas}, which in our normalized set of parameters corresponds to $\alpha=-1.178, \beta=0.01125$. In previous literature, this reference case has been shown to capture experimental results \cite{wensink12pnas}. As the value of $\alpha$ is decreased, the energy injection into the system becomes more intense and acts on a wider range of scales. As a result the energy spectra show an increased broadband excitation. Due to the inverse energy transfer the spectral peak gradually shifts from the most unstable wave number to smaller wave numbers, indicating the emergence of larger-scale flow structures. All of these trends are captured accurately by EDQNM without further adjustments. The EDQNM theory therefore extends the low-wave-number theory developed in \cite{bratanov15pnas} to the full range of scales. With the full energy spectra at hand, correlation functions can be computed in a straightforward manner. The results are shown in the right panel of Fig.~\ref{fig:edqnmresults}. As the flow becomes increasingly turbulent, the correlation length increases. This can be understood from the previous observations in spectral space. Through the inverse energy transfer, larger-scale structures are excited leading to longer-range correlations. Again, EDQNM captures these observations accurately. These findings highlight the crucial impact of the nonlinear advection on the system and motivate the exploration of the dynamics in the parameter range of strong nonlinear advection.

\section{Turbulent Pattern Formation}

\begin{figure*}
  \includegraphics[width=1.0\textwidth]{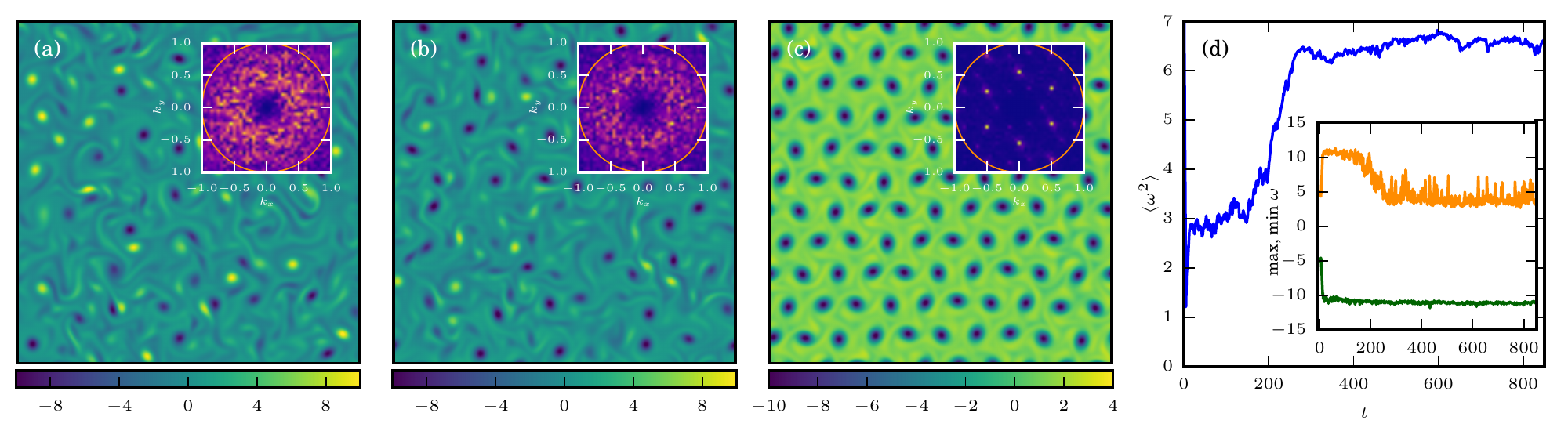}
  \caption{Emergence of hexagonal vortex lattice after a turbulent transient (simulation 4 in Table \ref{tab:simpara}). (a,b,c): Vorticity field after $t=20,150,850$. The insets show the two-dimensional vorticity spectra with the wave vectors corresponding to the most unstable wave number indicated by an orange circle. The inset (c) clearly shows six isolated peaks at $k_0 \approx 0.57$  which characterize the vortex lattice. For visualization purposes, these figures were obtained through a simulation on a smaller domain with half the domain length compared to Fig.~\ref{fig:dynamicalstates}. Note that the final vortex crystal state selects a sign of vorticity different from that of Fig.~\ref{fig:dynamicalstates}, exemplifying spontaneous symmetry breaking in this system. Panel (d) shows the evolution of the enstrophy, as well as the maximum and the minimum vorticity through the transient to the final quasi-stationary state.}
  \label{fig:vortexlattice}
\end{figure*}

Further increasing the strength of the nonlinear advection to $\lambda=7$ leads to a surprising new dynamical state emerging from a turbulent transient as visualized in Fig.~\ref{fig:vortexlattice}. From random initial conditions vortices arise, triggered by small-scale instabilities. Many vortices are screened by surrounding vorticity of opposite sign, reducing their Biot-Savart interaction. Some of them, however, form dipoles, which propagate rapidly through the flow. These dipoles contribute significantly to the turbulent dynamics. In the course of time, a spontaneous symmetry breaking occurs, such that one sign of vorticity prevails. As a result, less dipoles form and the dynamics stabilizes. Repeating the numerical experiment with different random initial conditions confirms that both vorticity signs are equally probable in this spontaneous symmetry breaking. By the continued emergence of vortices the system eventually crystallizes into a quasi-stationary hexagonal vortex lattice state. The wave number characterizing this turbulent pattern is significantly smaller than na\"ively expected based on the linear critical wave number $k_c=1$ in the classical pattern formation case. This can be explained as follows: as the turbulent pattern emerges out of a turbulent transient, there is an inverse transfer of energy feeding larger scales. As a result, the peak energy injection scale in Eq.~\eqref{eq:spectrumevo} (i.e.~the maximum of $2L(k,t)E(k,t)-T(k,t)$) shifts to smaller wave numbers during the transient, giving rise to larger-scale flow structures. Because $\int T(k,t) \mathrm{d}k=0$ by virtue of $T(k,t)$ being an energy transfer term, Eq.~\eqref{eq:spectrumevo} implies the constraint $\int L(k,t) E(k,t) \mathrm{d}k=0$ once the statistically stationary state with the vortex lattice is reached. Given the fact that the system forms a regular vortex pattern with a sharply localized spectrum around the lattice wave number, this constraint can only be satisfied if the lattice wave number $k_0$ is close to the zero-crossing of the effective linear term, i.e.~close to the wave number corresponding to the smallest neutral mode. For the current choice of parameters, this prediction yields $k_0 \approx 0.58$ in very good agreement with the numerical observation ($k_0 \approx 0.57$). To further confirm this prediction, we scanned the entire $\alpha$-range $[-0.95,-0.75]$ leading to stable vortex lattices, keeping all other parameters fixed. We observed a trend of the lattice wave number slowly increasing with $\alpha$, which is captured by the prediction to within ten percent (not shown). We conclude that this turbulent pattern formation selects the \emph{neutral} mode rather than the fastest growing linear mode. We stress that this mechanism profoundly differs from the Turing mechanism reported in \cite{grossmann14prl,grossmann15epje} due to the extended turbulent transient leading to the selection of the neutral mode.

It remains to explain the type of lattice. Nonlinear advection favors axisymmetric vortices. As these structures populate the domain over time, they form the densest possible packing consistent with this geometry, resulting in the hexagonal pattern. Unlike the case of classical pattern formation ($\lambda=0$), this vortex lattice is quasi-stationary with perturbations from weaker background turbulence. The most striking feature of this phenomenon is the long turbulent transient phase preceding the formation of the pattern, which lasts much longer than the typical lifetimes of the vortices in the turbulent phase. Furthermore, unlike classical pattern formation, the dominant length scale in the system is given by the neutral mode in the effective dispersion relation.

\section{Conclusions} 

The correlation functions and spectra of a minimal model for active turbulence developed in this paper establish a quantitative statistical theory of active turbulence. We adapted the EDQNM closure scheme for classical hydrodynamic turbulence to capture the linear driving and damping as well as the nonlinear energy transfer across scales along with nonlinear damping. For the range of investigated parameters, the theory has been found to accurately capture simulation results. It revealed that the spectral peak, associated with the typical size of turbulent flow structures, originates from the interplay of linear and nonlinear physics: energy is injected in a band of unstable modes which then cascades uphill before dissipated by linear and nonlinear damping terms. EDQNM therefore quantitatively captures the statistics of the collective behavior emerging in the continuum model Eq.~\eqref{eq:equationofmotion}. Having demonstrated the potential of methods from turbulence theory to capture disordered active matter states, we hope that our findings may spur further research. For instance, a generalization to active nematics might be an interesting direction for future research.

Further exploring the parameter space towards strong nonlinear advection, we find a highly ordered lattice state of dynamically self-organized vortices which emerges from an extensive turbulent transient. The inverse energy transfer of two-dimensional turbulence turns out to be a crucial ingredient in this turbulent pattern formation: the same mechanism leading to the spectral peak in the turbulent phase selects the neutral wave number in this turbulent pattern formation. While the potential importance of neutral modes has been pointed out in \cite{slomka17prf} based on kinematic considerations, our findings show that they are indeed dynamically relevant.

Regarding possible experimental realizations of the vortex lattice state reported here, we note that we observe it in a regime of strong nonlinear advection due to active stresses. Recent research has indicated that such a regime, in which the value of $\lambda$ is large, can be achieved by a microstate with strong polar interaction among the active particles \cite{reinken2018derivation}. Furthermore, we observe the vortex lattice in a parameter range (controlled by $\alpha$) of both large- and small-scale damping. Thus experiments involving active fluids with strong polar interactions and with substrate-mediated friction could potentially realize this novel ``turbulent pattern formation'' phenomenon.

Interestingly, the mechanism reported here shares similarity with quasicrystalline vortex lattices in drift-wave turbulence \cite{kukharkin95prl}, although their vortex pattern appear less stable than the ones reported here. Vortex crystals have also been observed in two-dimensional Navier-Stokes turbulence driven by a combination of deterministic and stochastic forcings \cite{jimenez07pof}, in truncated two-dimensional turbulence \cite{smithr1994finite}, in simulations of  quasi-geostrophic turbulence \cite{arbic2004effects} as well as in two-dimensional fluid films with polymer additives \cite{gupta2017melting}. Furthermore, vortex lattices have been predicted \cite{abrikosov57jpc} and observed \cite{essmann67pla} in superconductors. These observations in profoundly different physical systems point at the ostensibly universal occurrence of highly ordered states in strongly nonlinear regimes. The investigation of this phenomenon in generic systems which combine features of pattern formation with non-Lyapunov dynamics such as nonlinear advection appears as one exciting direction for future research.

\begin{acknowledgments}
This work was supported by the Max Planck Society. MJ gratefully acknowledges the financial support by the International Max Planck Research School ``Physics of Biological and Complex Systems'', G\"{o}ttingen.
\end{acknowledgments}

\bibliography{references.bib}
\pagebreak
\widetext
\begin{center}
\textbf{\large Supporting Information: Turbulence and turbulent pattern formation in a minimal model for active fluids}
\end{center}
\setcounter{equation}{0}
\setcounter{figure}{0}
\setcounter{table}{0}
\setcounter{page}{1}
\setcounter{section}{0}
\makeatletter
\renewcommand{\theequation}{S\arabic{equation}}
\renewcommand{\thefigure}{S\arabic{figure}}
\renewcommand{\thetable}{S\arabic{table}}

\section{Numerical Simulations}
\label{sisec:dns}

The numerical simulations are performed with a standard pseudospectral scheme for the vorticity formulation of Eq.~\eqref{eq:equationofmotion}:
\begin{equation}\label{eq:vorticityeq}
  \partial_t \omega + \lambda \bs u \cdot \bs \nabla \omega = - (1+\Delta)^2 \omega - \alpha \omega - \beta \,  \bs \nabla \times \left( \bs u^2\, \bs u \right) \, .
\end{equation}
Here $\omega(\bs x,t) = \bs \nabla \times \bs u(\bs x,t)$ is the pseudo-scalar vorticity. In turn, the velocity is obtained from the vorticity by Biot-Savart's law. An equation for the spatially constant velocity contribution $u_0$, which is not contained in the vorticity field, is integrated simultaneously. Time stepping is performed by means of a second-order Runge-Kutta scheme, in which the linear term is treated with an integrating factor. To account for the cubic nonlinearity, the pseudospectral scheme is fully dealiased with a $1/2$ dealiasing. Small-scale, low-amplitude random initial conditions are chosen for all simulations. The parameters for the various simulations are summarized in Table \ref{tab:simpara}.

\section{Classical pattern formation -- square lattice state}
\label{sisec:squarelattice}

For $\lambda=0$, Eq.~\eqref{eq:equationofmotion} follows a gradient dynamics constrained to the sub-space of incompressible velocity fields, $\partial_t \bs u = -\bs \nabla p - \delta {\cal L}[\bs u]/\delta \bs u$. Here, all terms except the pressure gradient can be combined into the Lyapunov functional (see also \cite{dunkel13njp,oza16epje})
\begin{equation}
{\cal L}[\bs u] = \int \!d\bs x \left[(\Delta \bs u + \bs u)^2/2  + \alpha \bs u^2/2+\beta \left( \bs u^2 \right)^2/4\right];
\end{equation}
 the pressure gradient term is a Lagrange multiplier to ensure $\bs \nabla \cdot \bs u=0$. As a result of the potential dynamics, a stationary pattern emerges; its wave number $k_c=1$ is straightforwardly computed by linear stability analysis.

\begin{figure}
  \includegraphics[width=0.48\textwidth]{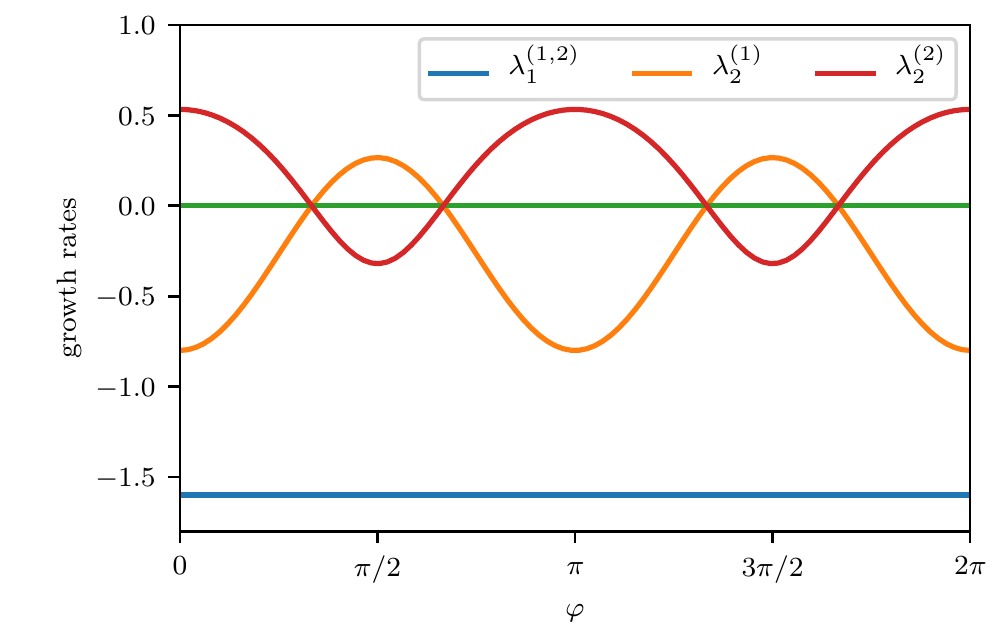}
  \caption{Growth rates of the linear stability analysis for $\alpha=-0.8$. The eigenvalues $\lambda_1^{(1,2)}$ correspond to the stable eigenvalues of the single- and two-stripe pattern, respectively. Starting from a single-stripe pattern, $\lambda_2^{(1)}$ indicates that a second stripe in a wave-number band around $\pi/2$ can be excited. The eigenvalue $\lambda_2^{(2)}$ shows that the square lattice state is linearly stable.}
\label{fig:linearstability}
\end{figure}

The pattern forming state can be conveniently analyzed in the vorticity formulation Eq.~\eqref{eq:vorticityeq}. Motivated by our numerical observations, we investigate a lattice state of the form
\begin{equation}
  \omega(\bs x,t) = \zeta_1(t) \exp[\mathrm{i} \bs k_1 \cdot \bs x] + \zeta_2(t) \exp[\mathrm{i} \bs k_2 \cdot \bs x] + \text{c.c.}
\end{equation}
with $|\bs k_1| = |\bs k_2| = k_c=1$, $\bs k_1 \cdot \bs k_2 / k_c^2= \cos \varphi$ and amplitudes $\zeta_1$ and $\zeta_2$, which we can choose as real due to translational invariance.
Combining this ansatz with the full nonlinear equations, amplitude equations can be straightforwardly derived, which in leading order take the form
\begin{align}
  \dot \zeta_1 &= -\alpha \zeta_1 - \frac{\beta}{k_c^2} \left( 3 \zeta_1^3 + 2\left[ 1+ 2 \cos^2 \varphi \right] \zeta_2^2 \zeta_1 \right) \\
  \dot \zeta_2 &= -\alpha \zeta_2 - \frac{\beta}{k_c^2} \left( 3 \zeta_2^3 + 2\left[ 1+ 2 \cos^2 \varphi \right] \zeta_1^2 \zeta_2 \right) \, .
\end{align}
These equations can be further analyzed by means of a linear stability analysis. The analysis shows that the ground state $\zeta_1=\zeta_2=0$ is linearly unstable for $\alpha<0$ with growth rates $\lambda^{(0)}_{1,2}=-\alpha$. For a single-stripe pattern with $\zeta_2=0$ the amplitude equations yield $\zeta_1 = \sqrt{-\alpha k_c^2/(3\beta)}$ as a stationary solution. A linear stability analysis with small perturbation of the single-stripe pattern yields growth rates $\lambda^{(1)}_1 = 2\alpha$ and $\lambda^{(1)}_2 = \alpha\left[ 4\cos^2\varphi -1 \right]/3$ (see Fig.~\ref{fig:linearstability}). As expected, small perturbations in the direction of the single stripe are damped for $\alpha<0$.  The emergence of a second stripe, however, is linearly unstable for a small wave number band around $\varphi = \pi/2$, which gives a first hint at the emergence of a square lattice. This can be further corroborated with a linear stability analysis of a lattice state with $\zeta_1=\zeta_2$, for which the stationary solution $\zeta_1 = \zeta_2 = \sqrt{\frac{-\alpha k_c^2}{\beta \left[ 5+4\cos^2\varphi \right]}}$ is readily obtained from the amplitude equations. Linear stability analysis, assuming small perturbations in both amplitudes, yields $\lambda^{(2)}_1 = 2\alpha$ and $\lambda^{(2)}_2=2\alpha\frac{1-4\cos^2\varphi}{5+4\cos^2\varphi}$. For $\alpha<0$ a range of lattice states is linearly stable with the maximum stability reached when $\varphi=\pi/2$ (see Fig.~\ref{fig:linearstability}). This analysis renders a clear picture of the emergence of square lattice states for $\alpha<0$: the single-stripe pattern is unstable with respect to the emergence of a two-stripe lattice with the maximum growth rate at $\varphi=\pi/2$. The resulting square lattice state with $\zeta_1=\zeta_2=\sqrt{-\alpha k_c^2/(5\beta)}$ then is linearly stable. Minimizing the Lyapunov functional for a square lattice with respect to the amplitude yields the same result.

\section{Active turbulence -- EDQNM closure}
\label{sisec:edqnm}

Developing a statistical theory for the turbulent phase of active fluids requires assumptions about the hierarchy of moments. Indeed, the equation for the covariance tensor Eq.~\eqref{eq:covarianceevo}, or equivalently, for the energy spectrum Eq.~\eqref{eq:spectrumevo}, is unclosed due to the presence of the higher-order velocity correlations stemming from the nonlinear terms. 

To set our theoretical development into context, we start with re-iterating the classical closure attempts in the context of the active turbulence equations. The classical closure theory is presented in much more detail in \cite{Orszag1974,lesieur2012turbulence,SagautBook}.
A Gaussian approximation is the simplest first choice to close the system, in particular in a random system like turbulence. Under this cumulant discard hypothesis, one can factorize higher-order moments in terms of corresponding second-order moments. This allows us to close the fourth-order term in Eq.~\eqref{eq:covarianceevo} as described in the main text. However, the third-order correlations in Eq.~\eqref{eq:spectrumevo} vanish under such a Gaussian approximation. The third-order correlations are responsible for the energy transfer between scales and hence are essential for the dynamics. A logical step towards closure is to write then the equations for the triple correlation, which in Fourier space take the form

\begin{align}
\left[\partial_t+\tilde{L}(k)+\tilde{L}(p)+\tilde{L}(q)\right]\langle \hat{u}({\bs k})\hat{u}({\bs p})\hat{u}({\bs q}) \rangle= F [\lambda\langle\hat{u}\hat{u}\hat{u}\hat{u}\rangle, \beta\langle\hat{u}\hat{u}\hat{u}\hat{u}\hat{u}\rangle] \, .
\end{align}
In favor of a lighter notation we write these equations rather schematically, suppressing tensorial notation. Here, $\tilde{L}(k)=(1-k^2)^2+\alpha$, and the functional $F$ captures the contributions due to the pressure term as well as the fourth- and the fifth-order correlations which appear due to the advective and cubic nonlinearities in Eq.~\eqref{eq:equationofmotion}, respectively. To close this system on the level of the quadruple and fifth-order correlations, one can now assume a Gaussian factorization of these higher-order moments as the next simplest closure. This eliminates the fifth-order correlations and the fourth-order correlation can now be written in terms of second-order correlations resulting in
 
 \begin{align}
\left[\partial_t+\tilde{L}(k)+\tilde{L}(p)+\tilde{L}(q)\right]\langle \hat{u}({\bs k})\hat{u}({\bs p})\hat{u}({\bs q}) \rangle= \lambda\Sigma\langle\hat{u}\hat{u}\rangle\langle\hat{u}\hat{u}\rangle
\end{align}
so that the Eq.~\eqref{eq:spectrumevo} for the energy spectrum $E(k)$ is now closed. This procedure is known as the quasi-normal approximation \cite{Millionschikov1941,Proudman1954}. This classical approximation for the energy transfer term has been shown to fail spectacularly for hydrodynamic turbulence already in the 1960s \cite{Ogura1963}, leading to a realizability problem by the development of negative energies, since the omission of the cumulants leads to an overprediction of the transfer term.

To remedy this shortcoming, more sophisticated manners of closure were proposed, in particular by Kraichnan \cite{KraichnanDIA,KraichnanLHDIA}, using renormalized perturbation theories. The simplest successful derivative of these theories is the eddy-damped quasi-normal Markovian model \cite{orszag1970analytical}. For an extensive account on the matter, we refer to \cite{Orszag1974,lesieur2012turbulence,SagautBook}. Here we adopt this framework to formulate a statistical theory for active turbulence.
The eddy-damped quasi-normal Markovian model generalizes the classical quasi-normal approximation by modeling the effect of the missing fourth-order cumulants as $\lambda(\langle\hat{u}\hat{u}\hat{u}\hat{u}\rangle - \Sigma\langle\hat{u}\hat{u}\rangle\langle\hat{u}\hat{u}\rangle)=-\mu_{kpq}\langle \hat{u}({\bs k})\hat{u}({\bs p})\hat{u}({\bs q}) \rangle$ where the damping term $\mu_{kpq}=\mu_k +\mu_p+\mu_q$ is defined through the contributions

\begin{equation}
 \mu_k = \lambda\gamma \left(\int_0^k s^2 E(s,t) ds\right)^{1/2} \, .
\end{equation}
Here, $\gamma$ is a free parameter which quantifies the strength of the eddy damping. We can then combine the linear terms to define $\eta_k=\mu_k + |\tilde{L}(k)|$ as the net damping. The damping of the triple correlation corresponds to the Lagrangian decorrelation of the Fourier modes \cite{Bos2013-1}, and both the positive and the negative linear terms will lead to an effective decorrelation. Consequently the effect of $\tilde{L}(k)$ in damping should be strictly positive, and hence we take the absolute value of $\tilde{L}(k)$. With this assumption, the evolution equation for the triple correlation can be written as

 \begin{align}
\left[\partial_t+ \eta_{kpq}\right]\langle \hat{u}({\bs k})\hat{u}({\bs p})\hat{u}({\bs q}) \rangle= \lambda\Sigma\langle\hat{u}\hat{u}\rangle\langle\hat{u}\hat{u}\rangle,
\end{align}
where $\eta_{kpq}=\eta_k+\eta_p+\eta_q$. If we neglect the time variation in $\mu_k$ and $\langle\hat{u}\hat{u}\rangle\langle\hat{u}\hat{u}\rangle$, the above expression can be integrated in time, resulting in the following expression for the triple correlation in terms of the energy spectrum:

 \begin{align}  \label{eq:t_k_eqn}
 \langle\hat{u}({\bs k})\hat{u}({\bs p})\hat{u}({\bs q}) \rangle(t)= \frac{1-e^{-\eta_{kpq}t}}{\eta_{kpq}}\lambda\Sigma\langle\hat{u}\hat{u}\rangle\langle\hat{u}\hat{u}\rangle .
\end{align}

For large time scales, $e^{-\eta_{kpq}t}$ can be neglected, and $1/\eta_{kpq}$ defines a characteristic time. This timescale is associated with the Lagrangian correlation time of the fluid particles (see for instance \cite{Bos2013-1} for a discussion). The second-order correlations are associated with the energy spectrum, hence Eq.~\eqref{eq:spectrumevo} and Eq.~\eqref{eq:t_k_eqn} together result in a closed set of equations for the evolution of the energy spectrum. Owing to the isotropy of the velocity field, $T(k)$ in Eq.~\eqref{eq:spectrumevo} can be calculated from $ \langle\hat{u}_l({\bs k})\hat{u}_m({\bs p})\hat{u}_n({{\bs q}})\rangle \equiv T_{lmn}({\bs k},{\bs p},{\bs q})$ (in full tensorial notation) as
\begin{align}
T(k)=\pi k P_{lmn}({\bs k})\int \text{Im}\left[T_{lmn}({\bs k},{\bs p},{\bs q})\right]d{\bs p}d{\bs q}  \, ,
\end{align}
where $P_{lmn}({\bs k})=k_n(\delta_{lm}-k_lk_m/k^2)+k_m(\delta_{ln}-k_lk_n/k^2)$ and Im stands for the imaginary part. The integration is performed over all triads  ${\bs k},{\bs p},{\bs q}$ where ${\bs k}+{\bs p}+{\bs q}=0$. The final expression for $T(k,t)$ can then be written as \cite{leith1971atmospheric}

\begin{align} \label{eq:edqnm_full}
T(k,t)=-\frac{4}{\pi}\iint_{\Delta}\frac{\lambda^2}{\eta_{kpq}} \, &\frac{xy-z+2z^3}{\sqrt{1-x^2}}\big[k^2pE(p,t)E(q,t)-kp^2E(q,t)E(k,t)\big]\frac{dpdq}{pq} \, .
\end{align}
Here $\Delta$ is a band in $p,q$-space so that the three wave numbers $k, p, q$ form the sides of a triangle. $x,y,z$ are the cosines of the
angles opposite to the sides $k,p,q$ in this triangle. Comparing Eq.~\eqref{eq:edqnm_full} with Eq.~\eqref{eq:edqnm}, we obtain $a(k,p,q)=-\frac{4}{\pi}\frac{xy-z+2z^3}{\sqrt{1-x^2}}\frac{k^2}{q}$ and $b(k,p,q)=\frac{4}{\pi}\frac{xy-z+2z^3}{\sqrt{1-x^2}}\frac{kp}{q}$.

To generate the results presented in the main text, this closed set of equations for the energy spectrum function is integrated numerically. Computations are carried out on a logarithmically spaced mesh on the interval $0.025\le k \le 25$ using 300 modes. All results are obtained, using $\gamma=0.55$, after the spectrum reached a steady state.
\end{document}